\documentclass[aps,amssymb,showpacs,twocolumn]{revtex4}
\usepackage{amssymb}
\usepackage{graphicx}
\usepackage{color}
\usepackage{textcomp}
\usepackage{ulem}
\begin{document}
	
	\title{ Inverse catalysis effect of quark anomalous magnetic moment to chiral restoration and deconfinement phase transitions}
	\author{Jie Mei}
	\author{Shijun Mao}%
	\email{maoshijun@mail.xjtu.edu.cn}
	\affiliation{School of Physics, Xi'an Jiaotong University, Xi'an, Shaanxi 710049, China}
	
	\begin{abstract}
		The effect of quark anomalous magnetic moment (AMM) to chiral restoration and deconfinement phase transitions under magnetic fields is investigated in a Pauli-Villars regularized Polyakov-extended Nambu–Jona-Lasinio model. A linear-in-$B$ term for quark anomalous magnetic moment is introduced to the Lagrangian density of our model, and it plays the role of inverse catalysis to the phase transitions. With fixed magnetic field, the critical temperature decreases with quark AMM. When fixing quark AMM, the critical temperature increases with magnetic field for a small quark AMM, but decreases with magnetic field for a large quark AMM. The critical temperature of chiral restoration and deconfinement phase transitions is determined by the two competing factors, the catalysis effect of magnetic field and inverse catalysis of quark anomalous magnetic moment.
	\end{abstract}
	
	\date{\today}
	\pacs{12.38.Aw, 11.30.Rd, 25.75.Nq, 12.39.-x}
	\maketitle
	
	Chiral symmetry restoration and quark deconfinement are the two most important quantum chromodynamics (QCD) phase transitions at finite temperature and baryon density. Motivated by the strong magnetic field in the core of compact stars and in the initial stage of relativistic heavy ion collisions, the study on QCD phase structure is recently extended to including external electromagnetic fields, see reviews~\cite{review1,review2,review3,review4,review5}. From recent lattice QCD simulations with a physical pion mass, while the chiral condensate is enhanced in vacuum, the critical temperature of the chiral restoration phase transition drops down with increasing magnetic field, which is the inverse magnetic catalysis effect~\cite{lattice1,lattice2,lattice3,lattice4,lattice5}. On the other hand, lattice simulations on the Polyakov loop also support the inverse magnetic catalysis for deconfinement phase transition, with a decreasing critical temperature as the magnetic field grows~\cite{lattice4,lattice5}. How to understand the inverse magnetic catalysis phenomena is still an open question~\cite{fukushima,mao,kamikado,bf1,bf12,bf13,bf2,bf3,bf4,bf5,bf8,bf9,bf11,db1,db2,db3,db5,db6,pnjl1,pnjl2,pnjl3,pnjl4,pqm,ferr1,ferr2,mhuang}.
	
	In the presence of a uniform external magnetic field $\bold B$, the energy dispersion of charged fermions takes the form $E=\sqrt{p_3^2+2eBl+m^2}$ due to the Landau quantization of the cyclotron frequencies characterized by the Landau level $l=0,1,2,...$~\cite{landau}. Due to this fermion dimension reduction, a magnetic catalysis effect on chiral symmetry breaking is expected in both vacuum and finite temperature in almost all model calculations at mean field level~\cite{mc1,mc2,mc3,review1,review2,review3,review4,review5}. Besides, the magnetic field also affects the radiative corrections of the fermion self-energy, which corresponds to the coupling between the field and the fermion anomalous magnetic moment (AMM)~\cite{amme1,amme2,amme3,amm0,amm1,amm2,amm3,amm4,amm5}. This gives rise to a new term $\frac{1}{2}a \sigma_{\mu \nu} F^{\mu \nu}$ in the Dirac Hamiltonian, with field tensor $F^{\mu \nu}$ and spin tensor $\sigma_{\mu \nu}=\frac{i}{2}[\gamma_\mu, \gamma_\nu]$, and the coefficient $a$ is identified as the fermion AMM, which is generally a function of magnetic field. The AMM term in the Hamiltonian changes the energy spectrum of fermions by removing the spin degeneracy and affects the properties of magnetized systems~\cite{ferr1,ferr2,mhuang,amm5,amm6,amm7}.
	
	In this paper, we will focus on the quark AMM effect on chiral restoration and deconfinement phase transitions. One of the models that describes well both the chiral restoration and deconfinement phase transitions is the Polyakov-extended Nambu--Jona-Lasinio (PNJL) model~\cite{pnjl5,pnjl6,pnjl7,pnjl8,pnjl9,pnjl10,pnjl12}. One problem in the (P)NJL model is the regularization. Since the model with contact interaction among quarks is nonrenormalizable, it requires a regularization scheme to avoid the divergent momentum integrations. By using the hard/soft cutoff regularization scheme, quark AMM effects on chiral restoration and deconfinement phase transitions have been studied in Ref~\cite{amm6,amm7}. When the external magnetic field is turned on, the quark energy becomes discrete and the phase space becomes anisotropy. In order to avoid nonphysical oscillations under magnetic field~\cite{amm6,amm7,reg1,reg3,reg4,mao,reg6}, we will apply a gauge covariant Pauli-Villars regularization scheme and investigate the chiral restoration and deconfinement phase transitions in this PNJL model. Different from the catalysis effect of magnetic field, the quark AMM plays the role of inverse catalysis to the critical temperature of phase transitions.
	
	The two-flavor PNJL model in external electromagnetic and gluon fields is defined through the Lagrangian density~\cite{pnjl5,pnjl6,pnjl7,pnjl8,pnjl9,pnjl10,pnjl12} in chiral limit,
	\begin{eqnarray}
		\mathcal{L}&=&\bar{\psi}(x)\left(i\gamma^{\mu}D_{\mu}+\frac{1}{2} a {\sigma}^{\mu\nu} F_{\mu\nu}\right)\psi(x)\\
		&&+\frac{G}{2}\left\{[\bar{\psi}(x)\psi(x)]^2+[\bar{\psi}(x)i\gamma_5 \vec{\tau}\psi(x)]^2 \right\}-\mathcal{U}(\Phi,\bar{\Phi}).\nonumber
		\label{lagrangian}
	\end{eqnarray}
	For the chiral section in the Lagrangian, the covariant derivative $D^\mu=\partial^\mu+i Q A^\mu-i {\cal A}^\mu$ couples quarks to the two external fields, the magnetic field ${\bf B}=\nabla\times{\bf A}$ and the temporal gluon field  ${\cal A}^\mu=\delta^\mu_0 {\cal A}^0$ with ${\cal A}^0=g{\cal A}^0_a \lambda_a/2=-i{\cal A}_4$ in Euclidean space. The gauge coupling $g$ is combined with the SU(3) gauge field ${\cal A}^0_a(x)$ to define ${\cal A}^\mu(x)$, $\lambda_a$ are the Gell-Mann matrices in color space, and $Q=diag(Q_u, Q_d)=diag(2e/3,-e/3)$ is the quark charge matrix in flavor space. The quark anomalous magnetic moment (AMM) is introduced by the term $\frac{1}{2} a {\sigma}_{\mu\nu} F^{\mu\nu}$, with spin tensor $\sigma_{\mu\nu}=\frac{i}{2}\left[\gamma_{\mu},\gamma_{\nu}\right]$, the Abel field strength tensor $F_{\mu\nu}=\partial_{[\mu,}A_{\nu]}$, and the quark AMM ${a}=Q { \kappa}$ and ${ \kappa}=diag({ \kappa}_u, { \kappa}_d)$ in flavor space. To simplify calculations, we assume a constant magnetic field ${\bf B}=(0, 0, B)$ along the $z$-axis and constant quark AMM $\kappa$ (or $a$). $G$ is the coupling constant in the scalar and pseudo-scalar channels. The order parameter to describe chiral restoration phase transition is the chiral condensate $\langle\bar\psi\psi\rangle$ or the dynamical quark mass $m=-G \langle \bar\psi\psi\rangle$.
	
	The Polyakov potential ${\cal U}(\Phi,{\bar \Phi})$ is related to the Z(3) center symmetry and simulates the deconfinement at finite temperature~\cite{pnjl6}
	\begin{eqnarray}
		\frac{\mathcal{U}}{T^{4}}=-\frac{b_{2}}{2}\bar{\Phi}\Phi-\frac{b_3}{6}\left(\bar{\Phi}^3+\Phi^3\right)+\frac{b_4}{4}\left(\bar{\Phi}\Phi\right)^2,
	\end{eqnarray}
	where the Polyakov loop is defined as $\Phi=\left({\text {Tr}}_c L \right)/N_c$, with $L({\bf x})={\cal P} \text {exp}[i \int^\beta_0 d \tau {\cal A}_4({\bf x},\tau)]= \text {exp}[i \beta {\cal A}_4 ]$ and $\beta=1/T$, the coefficient $b_2(t)=a_0+a_1 t+a_2 t^2+a_3 t^3$ with $t=T_0/T$ is temperature dependent, and the other coefficients $b_3$ and $b_4$ are constants. Polyakov loop $\Phi$ is considered as the order parameter to describe the deconfinement process, since it satisfies $\Phi \rightarrow 0$ in confined phase at low temperature and $\Phi \rightarrow 1$ in deconfined phase at high temperature~\cite{pnjl5,pnjl6,pnjl7,pnjl8,pnjl9,pnjl10,pnjl12}. Note that there is $\Phi=\bar\Phi$ at vanishing baryon density.
	
	Taking mean field approximation, the thermodynamic potential contains the mean field part and quark part
	\begin{eqnarray}
		\label{omega1}
		\Omega_{\text {mf}} &=&{\cal U}(\Phi)+ \frac{m^2}{2 G}+\Omega_q,\\
		\Omega_q &=& - \sum_{f,n,s} \int \frac{d p_z}{2\pi} \frac{|Q_f B|}{2\pi} \big[3E_f\nonumber\\
		&+& 2T\ln\left(1+3\Phi e^{-\beta E_f}+3{ \Phi}e^{-2\beta E_f}+e^{-3\beta E_f}\right)\big],\nonumber
	\end{eqnarray}
	with quark energy
	\begin{eqnarray}
		E_f=\sqrt{p^2_z+\left(\sqrt{\left(2 n+1-s \xi_f \right) |Q_f B|+m^2}-s \kappa_f Q_f B \right)^2 } \nonumber
		\label{ef}
	\end{eqnarray}
	for flavor $f$, longitudinal momentum $p_z$, Landau level $n$, spin $s$ and sign factor $\xi_f={\text {sgn}}(Q_f B)$.
	
	Note that the quark energy dispersion $E_f$ is straightforwardly derived from the modified Dirac equation $(\gamma\cdot\Pi-m+\frac{1}{2}{a}\sigma^{\mu\nu}F_{\mu\nu})\psi(x)=0$ with $\Pi_\mu=i \partial_\mu-QA_\mu$ by using the Ritus eigenfunction method~\cite{ritus1,ritus2,ritus3}. In the lowest-Landau-level approximation, we have quark energy $E_f=\sqrt{p^2_z+m_{\text {eff}}^2 }$, with effective quark mass $m_{\text {eff}}=m- \kappa_f |Q_f B|$. The quark AMM affects the system through the contribution to the effective quark mass. With vanishing quark AMM $\kappa_f=0$, the magnetic field causes the catalysis effect to the quark mass $m=m_{\text {eff}}$, and thus to the critical temperature of chiral restoration phase transition~\cite{mc1,mc2,mc3,review1,review2,review3,review4,review5}. For positive $\kappa_f$, the effective quark mass $m_{\text {eff}}$ will be smaller than quark mass $m$, which indicates that the quark AMM might induce an inverse catalysis effect to the chiral restoration phase transition. Therefore, the competition between the inverse catalysis effect of quark AMM and the catalysis effect of magnetic field determines the critical temperature of phase transition.
	
	The ground state is determined by minimizing the thermodynamic potential, $\partial\Omega_{\text {mf}}/\partial m=0$ and $\partial\Omega_{\text {mf}}/\partial \Phi=0$, which leads to the two coupled gap equations for the two order parameters $m$ and $\Phi$,
	\begin{eqnarray}
		\label{gap1}
		&& m\left( \frac{1}{2G}+\frac{\partial \Omega_q}{\partial m^2}\right)=0,\\
		\label{gap2}
		&& \frac{\partial {\cal U}}{\partial \Phi}+\frac{\partial \Omega_q}{\partial \Phi}=0.
	\end{eqnarray}
	From Eq.(\ref{gap1}), we can always find a solution $m=0$. The chiral restoration phase transition happens when non-vanishing quark mass $m$ turns into zero. At this time, the two coupled gap equations Eq.(\ref{gap1}) and Eq.(\ref{gap2}) become decoupled. In the chiral restoration phase with $m=0$, we only need to solve Polyakov loop $\Phi$ from Eq.(\ref{gap2}). Thus, we obtain the same critical temperature for chiral restoration and deconfinement phase transitions in chiral limit~\cite{mao}.
	
	Because of the contact interaction among quarks, NJL models are nonrenormalizable, and it is necessary to introduce a regularization scheme to remove the ultraviolet divergence in momentum integrations. The magnetic field does not cause extra ultraviolet divergence but introduces discrete Landau levels and anisotropy in momentum space. The usually used hard/soft momentum cutoff regularization schemes do not work well in magnetic field, since the momentum cutoff together with the discrete Landau levels will cause some nonphysical results~\cite{amm6,amm7,reg1,reg3,reg4,mao,reg6,reg7}, such as the oscillations of chiral condensate and critical temperature and density, tachyonic pion mass, and the breaking of law of causality for Goldstone mode. In this work, we take into account the gauge covariant Pauli-Villars regularization scheme~\cite{mao,rev6}, where the quark momentum runs formally from zero to infinity, and the nonphysical results are cured~\cite{reg4,reg6}. Under the Pauli-Villars scheme, one introduces the regularized quark energy $E_{f,i}=\sqrt{p^2_z+M_{\text {eff}}^2+a_i \Lambda^2}$ with $M_{\text {eff}}=\sqrt{\left(2 n+1-s \xi_f \right) |Q_f B|+m^2}-s \kappa_f Q_f B$ and the summation and integration $\sum_n\int dp_z/(2\pi)F(E_f)$ is replaced by $\sum_n\int dp_z/(2\pi)\sum_{i=0}^N c_iF(E_{f,i})$. The coefficients $a_i$ and $c_i$ are determined by constraints $a_0=0$, $c_0=1$, and $\sum_{i=0}^N c_i\left(m^2+a_i\Lambda^2\right)^{L}=0$ for $L=0,1,\cdots N-1$. In chiral limit there are two parameters, the quark coupling constant $G$ and Pauli-Villars mass parameter $\Lambda$. By fitting the pion decay constant $f_\pi=93$ MeV and chiral condensate $\langle \bar \psi \psi \rangle =(-250\ \text{MeV})^3$ in vacuum, the two parameters are fixed to be $G=7.04$ GeV$^{-2}$ and $\Lambda=1127$ MeV in Pauli-Villars scheme with number of regulated quark masses $N=3$. On the Polyakov potential, its temperature dependence is from the lattice simulation, and the parameters are chosen as~\cite{pnjl6} $a_0=6.75$, $a_1=-1.95$, $a_2=2.625$, $a_3=-7.44$, $b_3=0.75$, $b_4=7.5$ and $T_0=270$ MeV. To evaluate the effect of quark AMM, we consider two typical sets of parameters $\kappa$,
	\begin{eqnarray}
		\kappa_u^{(1)}&=&0.00995\ {\text {GeV}}^{-1},\\
		\kappa_d^{(1)}&=&0.07975\ {\text {GeV}}^{-1}, \nonumber
	\end{eqnarray}
	and
	\begin{eqnarray}
		\kappa_u^{(2)}&=&0.29016\ {\text {GeV}}^{-1},\\
		\kappa_d^{(2)}&=&0.35986\ {\text {GeV}}^{-1}, \nonumber
	\end{eqnarray}
	which are phenomenologically determined by fitting the nucleon magnetic moments~\cite{amm6,amm7,amm2}.
	In our numerical calculations, we consider all the Landau levels and longitudinal momenta.
	
	\begin{figure}[hb]
		\centering
		\includegraphics[width=9.2cm]{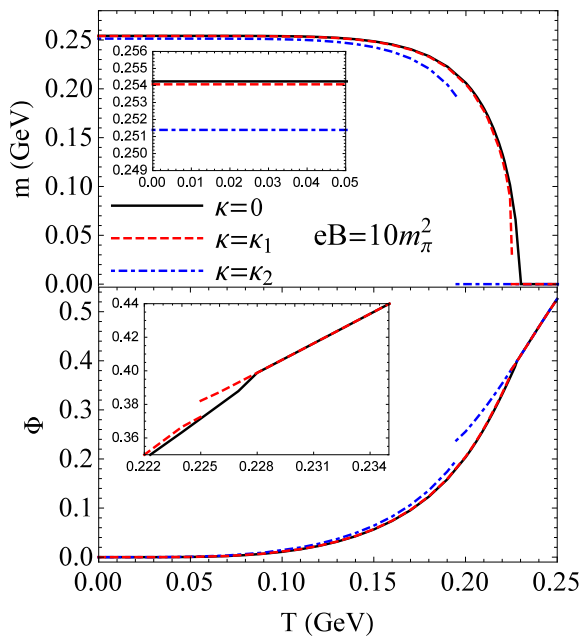}
		\caption{ The quark mass $m$ and Polyakov loop $\Phi$ as functions of temperature $T$ with fixed external magnetic field ($eB=10m_{\pi}^2$ and $m_{\pi}=134$ MeV) and different sets of quark AMM, ${\kappa=0}$ (black solid lines), ${\kappa=\kappa_1}$ (red dashed lines), ${\kappa=\kappa_2}$ (blue dot-dashed lines). }
		\label{fig1}
	\end{figure}
	We firstly discuss the quark AMM effect on chiral restoration and deconfinement phase transitions, and plot in Fig.\ref{fig1} the quark mass $m$ and Polyakov loop $\Phi$ as functions of temperature with different sets of quark AMM and fixed external magnetic field ($eB=10m_{\pi}^2$ and $m_{\pi}=134$ MeV), where the black solid lines are for vanishing quark AMM $\kappa=0$, red dashed lines for small quark AMM $\kappa=\kappa_1$ and blue dot-dashed lines for large quark AMM $\kappa=\kappa_2$. Temperature melts the chiral condensate and causes the chiral restoration phase transition. The critical temperature $T_c$, determined by zero quark mass, decreases with increasing $\kappa$, which indicates the inverse catalysis effect of quark AMM. For vanishing quark AMM $\kappa=0$, we observe a second order chiral restoration phase transition with quark mass continuously approaching to zero. For nonvanishing quark AMM $(\kappa_1,\ \kappa_2)$, the chiral restoration turns into a first order phase transition, and quark mass jumps to zero. In chiral breaking phase, the quark mass $m$ decreases with quark AMM $\kappa$. However, the mass jump at $T_c$ increases with $\kappa$. The Polyakov loop $\Phi$ in the lower panel grows with temperature from zero to nonzero value, due to the deconfinement process. The critical temperature of deconfinement phase transition is defined by the fastest change of Polyakov loop $\Phi$. We obtain the same critical temperature for deconfinement and chiral restoration phase transitions. Since at critical temperature $T_c$, we have $m=0$, and the two coupled gap equations Eq.(\ref{gap1}) and Eq.(\ref{gap2}) become decoupled. For $T \geq T_c$, we separately solve quark mass $m=0$ from gap equation Eq.(\ref{gap1}) and Polyakov loop $\Phi$ from Eq.(\ref{gap2}). For vanishing quark AMM $\kappa=0$, a slower increase of $\Phi$ is observed at $T > T_c$. For nonvanishing quark AMM $(\kappa_1,\ \kappa_2)$, Polyakov loop $\Phi$ shows a finite jump at $T_c$, indicating a first order deconfinement phase transition. When approaching the critical point $T_c$, $\Phi$ shows apparant dependence on quark AMM $\kappa$. With larger $\kappa$, Polyakov loop $\Phi$ and its jump at $T_c$ are larger.
	
	\begin{figure}[hb]
		\includegraphics[width=9.4cm]{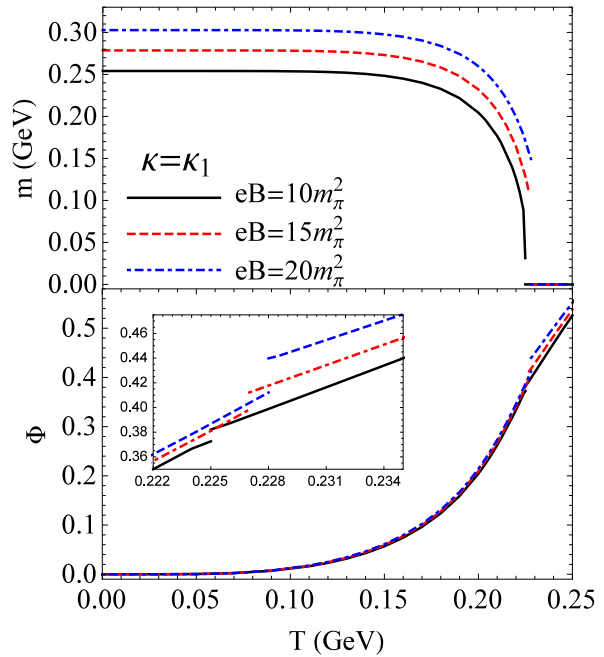}\\
		\includegraphics[width=9.4cm]{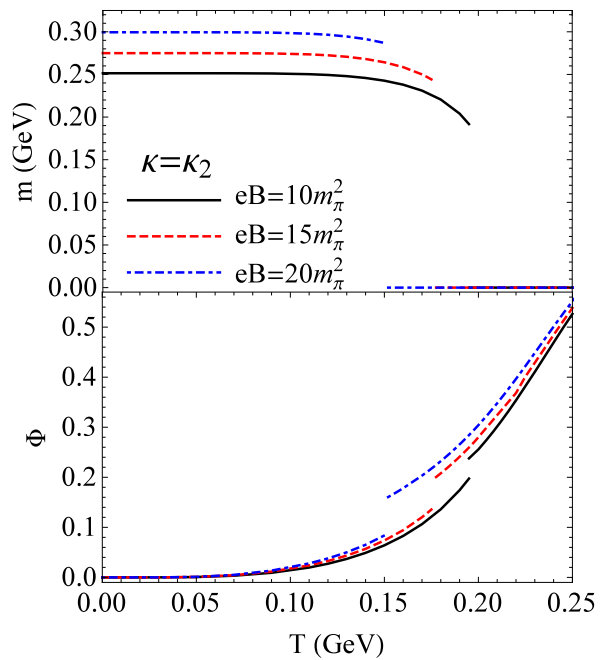}
		\caption{The quark mass $m$ and Polyakov loop $\Phi$ as functions of temperature with fixed quark AMM $\kappa=\kappa_1$ (upper panel) and $\kappa=\kappa_2$ (lower panel) and different magnetic field, $eB=10m^2_\pi$ (black solid lines), $eB=15m^2_\pi$ (red dashed lines), and $eB=20m^2_\pi$ (blue dot-dashed lines).}
		\label{fig3}
	\end{figure}
	Fig.\ref{fig3} focuses on the magnetic field effect to the chiral restoration and deconfinement phase transitions with finite quark AMM. With nonvanishing quark AMM $(\kappa_1,\ \kappa_2)$, the chiral restoration and deconfinement are both first order phase transitions under finite magnetic field, with quark mass and Polyakov loop jumping at the same critical temperature $T_c$. For a small quark AMM $\kappa=\kappa_1$ shown in the upper panel, the quark mass increases with magnetic field in the chiral breaking phase, and the Polyakov loop increases with magnetic field in the whole temperature region. The critical temperature $T_c$ for chiral restoration and deconfinement phase transitions increases with magnetic field. This magnetic catalysis effect is similar as the case of vanishing quark AMM $\kappa=0$~\cite{mc1,mc2,mc3,review1,review2,review3,review4,review5}. However, when we consider a large quark AMM $\kappa=\kappa_2$, although the quark mass and Polyakov loop increases with magnetic field, the critical temperature $T_c$ for chiral restoration and deconfinement phase transitions decreases with magnetic field, which shows the inverse magnetic catalysis. With the increase of quark AMM, the magnetic catalysis effect turns into the inverse magnetic catalysis effect, and the critical values of quark AMM are $\kappa_u^{(c)}=0.05072\ {\text {GeV}}^{-1}$ and $\kappa_d^{(c)}=0.12422\ {\text {GeV}}^{-1}$. The physics can be understood as follows. The critical temperature $T_c(\kappa, {\bold B})$ is determined by the two competing factors, the inverse catalysis effect of quark AMM $\kappa$ and catalysis effect of magnetic field ${\bold B}$. Therefore, with small quark AMM $\kappa=\kappa_1$, the catalysis effect of magnetic field dominates and the critical temperature increases with magnetic field, but with large quark AMM $\kappa=\kappa_2$, the inverse catalysis effect of quark AMM dominates and the critical temperature decreases with magnetic field.
	
	\begin{figure}[hb]
		\includegraphics[width=7cm]{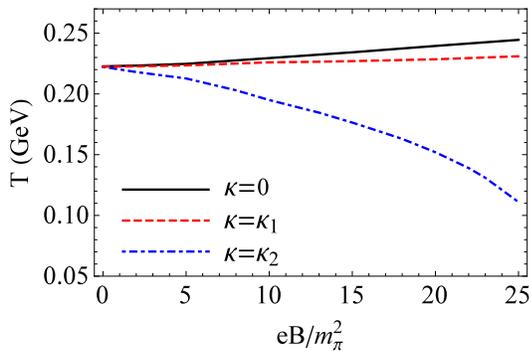}
		\caption{The phase diagram of chiral restoration and deconfinement phase transitions in $eB-T$ plane for different sets of ${\kappa}$. In chiral limit, the two critical temperatures coincide.}
		\label{fig5}
	\end{figure}
	Phase diagram in $eB-T$ plane is summarized in Fig.\ref{fig5}, where the phase transition lines of chiral restoration and quark deconfinement are depicted with fixed quark AMM $\kappa=0,\ \kappa_1,\ \kappa_2$. In low (high) temperature region, the chiral symmetry is spontaneously broken (restored) and quarks are confined (deconfined). The critical temperatures $T_c$ of chiral restoration and deconfinement phase transitions coincide in chiral limit. The quark AMM plays the role of inverse catalysis to the critical temperature, which leads to a lower $T_c$, and the magnetic field takes the role of catalysis, which leads to a higher $T_c$. The competition between them determines the phase structure. When fixing magnetic field, the critical temperature decreases with quark AMM. The stronger the magnetic field is, the faster the critical temperature decreases with $\kappa$. With fixed quark AMM $\kappa=0$, the critical temperature increases with magnetic field. For a small quark AMM $\kappa=\kappa_1$, critical temperature still increases with magnetic field, but with a slower increase ratio compared with vanishing quark AMM $\kappa=0$ case. For a large quark AMM $\kappa=\kappa_2$, the critical temperature decreases with magnetic field. With vanishing quark AMM $(\kappa=0)$, the chiral restoration and deconfinement phase transitions are of second order. But, with nonvanishing quark AMM $(\kappa_1,\ \kappa_2)$, the phase transitions are of first order.
	
	The effect of quark anomalous magnetic moment (AMM) to chiral restoration and deconfinement phase transitions under magnetic fields is investigated in frame of a Pauli-Villars regularized PNJL model. Different from the catalysis effect of magnetic field, quark AMM plays the role of inverse catalysis to the phase transitions. With fixed magnetic field, the critical temperature decreases with quark AMM. For a large enough quark AMM, the critical temperature decreases with magnetic field, while for vanishing or small quark AMM, critical temperature increases with magnetic field. The chiral restoration and quark deconfinement becomes first order phase transitions with nonvanishing quark AMM, and the critical temperatures $T_c$ coincide with each other.
	
    From the comparison with the cutoff regularization~\cite{amm6,amm7}, the covariant Pauli-Villars regularization plays an important role in avoiding the oscillations. To see if the covariance controls the calculation, one needs to compare with other regularization schemes, such as the magnetic field independent regularization (MFIR)~\cite{mfir1,mfir3,reg4}, which successfully separates the magnetic contribution from the vacuum thermodynamic potential and also efficiently avoids the nonphysical oscillations. Since the quark AMM $\kappa_f$ reduces the effective quark mass with $m_{\text {eff}}=m- \kappa_f |Q_f B|$, the quark AMM induced inverse catalysis effect will be independent of the regularization scheme.
	
	\noindent {\bf Acknowledgement:}
	The work is supported by the NSFC Grant No. 11775165 and Fundamental Research Funds for the Central Universities.

\end{document}